# Practices to Improve Teamwork in Software Development During the COVID-19 Pandemic: An Ethnographic Study

Ronnie E. de Souza Santos
Faculty of Computer Science
Dalhousie University
Halifax, NS, Canada

Paul Ralph
Faculty of Computer Science
Dalhousie University
Halifax, NS, Canada

## ABSTRACT

**Context**. Due to the COVID-19 pandemic, software professionals had to abruptly shift to ostensibly temporary home offices, which affected teamwork in several ways. **Goal**. This study aims to explore how these professionals coped with remote work during the pandemic and to identify practices that supported the team activities. **Method**. Ethnographic methods, including participant observation and qualitative data analysis, were used. **Results**. Three practices were created by the observed team to improve their engagement: costume meeting, second-language day, and project happy hour. These practices appear to increase individual involvement, improve team cohesion, reduce monotony, and create opportunities for knowledge acquisition. **Conclusions**. The three observed practices may help remote software teams cope with adversity. More research is needed to determine if these practices work in other settings, including remote-first and hybrid remote / on-site teams, post-pandemic.

## 1 Introduction

The COVID-19 crisis affected many aspects of our society, forcing us to adjust the way we live, work, and interact with others. In particular, the pandemic triggered a shift to remote work for millions of professionals around the globe [1][2][3].

Studying this unforeseen scenario is important because this crisis might produce long-term changes in work-patterns across many industries [4][5] and because it can give us a window into future long-term disasters and crises like climate change. In the context of software engineering, for instance, even though many IT professionals are used to distributed software development, there are characteristics of teamwork (especially human-related factors) that vary depending on where and how the software process is conducted [6], especially considering the reality of working during a worldwide crisis.

The human aspects of software development have been addressed in many studies over the years because the software development process strongly depends on human-based activities [7]. However, recent pandemic-induced upheaval casts doubt on much conventional wisdom and triggers novel phenomena. For instance, when software professionals had to abruptly shift into ostensibly temporary home offices in the beginning of the pandemic, practitioners were forced to develop strategies to support team activities [4][5].

Now, while many workers are returning to offices, more software professionals want to continue working from home, motivated by more flexibility and improved work-life balance [12][13]. Thus, hybrid teams, in which some members work on-site while others work remotely, are expected to become more common [6][28][32].

In this study, we followed an ethnographic methodological approach to accompany the routine of two software teams working remotely during the pandemic, and observed practices created to improve the engagement among team members. Such practices can help us to answer the following research question: *(RQ) how did software professionals and software companies improve teamwork during the COVID-19 pandemic?*

From this introduction, the present study is organized as follows. In Section 2, we discuss the use of ethnography as a research method. In section 3, we present how we designed and conduced our ethnography. In section 4, we present our main findings, which are discussed in section 5. Finally, in section 6, we present our conclusions.

## 2 Ethnographic Studies

Ethnography is a qualitative research method oriented by the thorough observation of individuals and their interaction [11]. Ethnographies differ from other scientific methods in its focus, since it seeks to understand what is and what is not relevant and important for the study participants, based on what these individuals experienced in real life [11]. Ethnography originated from anthropology and social sciences; however, its techniques has been successfully applied in many other areas over the years, allowing the study of cultural and social aspects of organizations and groups [15].

Depending on the place where the research is conducted, an ethnography can be classified as [14]:

- *Classic:* the traditional form of ethnography, originating from anthropology, based on in-depth and holistic observation, followed by empirical description of experiences [14].
- *Multi-site:* an evolution of classic ethnography focusing on interdisciplinary work (not only the study of a culture), such as science and technology [16].
- *Netnography:* the study of online communities (e.g., forums and social media) by analyzing posts, messages, and interactions among members [17].
- *Digital ethnography:* exploring the potential of computer-based storytelling and the perception of culture that flows across online environments, through virtual communication content, both graphic (e.g., images, emoticons, videos) and verbal (posts and discussions) [18].

Ethnography can also be categorized based on the perspective targeted by the data collection [19]:

- *Life history* is based on the individual viewpoint that participants hold about life events.
- *Memoir* is a well-structured report of everything that the researcher experienced while observing in the field.
- *Narrative* is an integration of life history and memoir: the researcher relies on observations of daily events that affect participants' lives, including their own perceptions about individual reactions.
- *Auto-ethnography:* the researcher is the participant and writes findings about their own experience.

Finally, ethnography is based on the researcher's immersion into the participants' environment, and there are two main categories of immersion [20]:

- *Participant observation*: researchers join in with the participants' activities, allowing them to experience and feel the reality first-hand.
- *Non-participant observation*: researchers observe actions and behaviors, but do not practice them during the research.

In software engineering, the analytical power of ethnography has been overlooked by researchers [15]. However, ethnography can produce detailed explanation of how software development occurs, considering aspects of teamwork and highlight the reasons and whys behind professional's actions towards practices, processes, and methods. Therefore, in this study, we advocate the relevance of applying an ethnographic approach to understand the dynamics of the work in software organizations, considering human aspects and the process of working from home during the pandemic.

## 3  Method

The present study is part of a broad research initiative focused on understanding the effects of the pandemic in the software industry, and encapsulating that understanding in parsimonious theories (e.g. [6]). However, some of our findings do not contribute directly to theory, but can be used directly by practitioners.

In this article, therefore, we present and discuss practices that were created to support teamwork and team engagement during the pandemic period, and that can potentially be used by hybrid teams in the period post-pandemic. These practices were identified in our general research when we applied techniques of ethnography to dive into the reality of software professionals that were working remotely during the pandemic.

We combined elements of *netnography* and *digital ethnography*, since the data was collected through digital environments (e.g., forums and discussion groups), using different sources of information both digital (e.g., posts, discussions, imagens) and traditional (observation and questionnaire). We adopt a *narrative* perspective to identify the practices implemented by the team through the observation of events (e.g., virtual meetings) and the perspective of professionals (questionnaire). In addition, elements of *auto-ethnography* and *participant observation* were applied, since the first author of this paper worked with the teams during data collection. As for the methodological phases that guided this research, we established the steps to conduct our ethnographic study following the method applied in previous ethnographies successfully conducted in software engineering [25][26][27][34], as presented below.

### 3.1  Entering the Field

We conducted our study on a mature software company that adopted an emergency work arrangement to move software professionals from the office to their homes at the beginning of the pandemic. This company specializes in on-demand software development for external clients and has over 1000 employees, of which about 70% work directly with software development (i.e., programmers, QAs, designers, analysts, etc.). The company has approximately 50 projects running simultaneously in several business areas. The first author was one of the employees affected by the emergency pandemic plan and for 18 months observed the dynamics of his team (e.g., rotation of team members between projects, arrival of newcomers, establishment of new practices) while working from home. More details regarding the team are presented in Section 4.

### 3.2  Data Collection

Data collection involved multiple techniques including participative observation in virtual environments, field notes, posts and discussions, and opinions obtained from a questionnaire. We consistently collected data focusing on the work of two software projects, the introduction of practices to support the team, and its effects on participants.

Our observations followed the fundamentals of netnography and digital ethnography, since we were focused on collecting messages, posts, and media about the practices. In addition, the participant observation happened during virtual meetings of projects (dailies, retrospectives) by taking notes about the interactions within the team. We also used an open-ended questionnaire about each practice introduced in the process, asking participants to provide their perspectives on the effects of each practice on their work and their team. The questions were the following: (a) *How did [practice] affected your work while you were working from home?* (b) *How did [practice] affected the team while you were working from home?*

### 3.3 Data Analysis

Data analysis in an ethnography study tends to be challenging due to the large amount of qualitative data collected in many formats (e.g., text, audio, images, notes) [11][14]. Following previous research and to make the data analysis feasible, we converted all data into text to enable narrative analysis [31]. Hence, digital data and media, such as audio, video, images, emoticons, among others, were converted into a textual description and combined with field notes from the participant observation process. The same analysis was applied to the answers of participants to the open-ended questions. Finally, we applied narrative analysis [31] to consolidate, reduce, and interpret data obtained from the various sources, and make sense of them.

## 4 Results

When the emergency measures started in March 2020, we started following the everyday events of one team (Project A). This project focused on developing and maintaining a mobile application for an international company. The team was composed of two developers, one quality assurance analyst (QA), one tech leader and one project manager.

In June 2020, a new project (Project B) was created by splitting Project A in two, due to its expanding scope. Project B entailed developing a desktop application for the same client. The team comprised the same software QA, tech leader and manager, plus two developers (one from project A and one new employee). Both projects A and B used a software process that included some Scrum practices, such as sprints and daily meetings.

In December 2020, both teams started to be drastically modified. A rotation of professionals brought a new team leader and a new manager to work in both projects. In addition, a new QA (also shared by both teams) joined the project, as well as a new developer for Project A and two new developers for Project B. In January and February 2021, the levels of engagement among team members were alarmingly low, as the team members had not created connections among them. Projects A and B appeared detached, even though they shared the same client, several team members, and some resources and features.

The new management began modifying the software development process to improve engagement among professionals. These practices included changes in the daily meetings, planning, and the establishment of retrospective meetings. However, since these practices are more related to the process and less related to human interaction, the levels of engagement were still not satisfactory. On the other hand, there was still a gap between Projects A and B, as both projects have their own events.

The idea to make meetings less formal and create opportunities for team members to get to know each other better, arose from the team members themselves, as an attempt to improve team engagement, cohesion, and the relationship between the two projects. To this end, the teams adopted three specific practices:

*Costume Meeting* - team members turn on their video feeds and participate in daily meetings using an improvised costume. The costume meeting was created to improve the relationship among teammates. Mainly, the strategy to turn on the camera during the meeting helped professionals get to know each other better. We observed that the costumes made individuals more interested in knowing their teammates' behaviors, resulting in improved levels of interaction among individuals. In addition, professionals reported that this practice reduced monotony and boredom associated to remote work.

*Second Language (English) Day*: team members speak only the second language used by the team during the whole day (e.g., in Project A and B, the official second language was English). The [Second Language] Day ended up being a practice less related to increasing interaction and more related to learning processes. We observed that professionals were extremely motivated by the learning opportunities created by the collaboration among teammates.

*Project Happy Hour:* the two teams merge for an online happy hour meeting where speaking about work is not allowed. Participants would turn on their video, all using the same background (simulating being at the same caffe, bar, party, etc.). Participants' conversation and activities (music, cinema, games, among others) focus on topics that the participants like from outside the work. The Project Happy Hour was the most successful practice in increasing team engagement. It not only helped to make connections that support team cohesion, but it also reduced the isolation resulting from remote work. After this practice, we constantly observed increasing levels of interaction among teammates. We also observed that professionals who have left the team (and the company) continue to participate of these activities. This is relevant because the expected demands for flexibility and remote work tend to intensify after the pandemic. Therefore, software companies need to create strategies to reduce turnover [28] and bring competent professionals to their teams. As the Project Happy Hour practice seems to support long-term connections, companies can strategize ways to explore these networks.

## 5 Discussions

Although these practices were not designed following any formal theory, they are share elements used in gamification [8] [9*]*. All three practices had: (1) clear *goals*—making the workplace fun and helping professionals to relax; and (2) included *rules* participants must follow, "as shown in Table 1". Thus, as we analyzed these practices and the rules, we associate them with aspects of gamification. Gamification is the process of making something that isn't a game more game-like, typically by adding elements associated with games [8]. Gamification has a spotty track record but enriching activities with game elements can, sometimes, make them more fun and engaging [9].

Through gamification we can define objectives and rules to guide individuals' actions in a non-game context, allowing the emergence of interactions based on personality and behaviors [21]. The list of game elements that can be applied in *non-game* contexts is vast as such elements are highly flexible. However, they can be grouped in categories related to individual performance, environmental properties, social interactions, individual sensations, and the use of fictional elements [22].

Gamification has the potential to improve motivation of software professionals [23]. Over the years the gamification elements more presented in software processes are related to performance, such as, points, badges, and acknowledgment (awards). However, the literature also presents evidence on the use of ecological elements (i.e., luck, paths) and fictional elements [10].

The costume meeting carries gamification elements as it takes advantage of environmental properties (e.g., visual features [22]) to create interaction among team members (as if they were game players). Second-language day employs elements of gamification as it applies social aspects of games (cooperation and social pressure [22]) while it is played around one main rule: everybody needs to speak only in the second language. Meanwhile, the project happy-hour practice involved elements of storytelling [22].

**Table 1: Techniques Supporting Remote Work in the Software Industry**

| Technique | Participants' Quotes | Game Element | Effect | Lessons Learned |
|---|---|---|---|---|
| Costume Meeting | "In the beginning of the pandemic was very interesting, because we didn't know several people on the team, so this practice brought us together". (P05). <br><br> "It became an entertaining moment to reduce job monotony." (P01) <br><br> "Besides making the team more relaxed, I believe it also contributed to increase team cohesion." (P06) | *Novelty:* use of the updates within the environment by adding new information and content [22] (e.g., different costumes every week). <br><br> *Sensation*: use of sound and visual stimulation to improve experiences [22] (e.g., visualization of teammates workstation). | *Positive:* It increases the interaction among team members and supports team engagement. <br><br> *Negative:* It can become demanding over time. | The positive effect of this practice tends to weaken over time, since it requires high levels of creativity (e.g., for costumes). We suggest making the use of costumes optional and replaceable by different accessories (e.g., mugs, puppets, collection items, etc.), or virtual visual elements (e.g., photos, backgrounds, etc.). We also suggest reducing the frequency of the practice (e.g., biweekly, monthly, or special occasions). |
| [Second Language] Day | "This practice helped me to be more confident to talk to the client whenever I need." (P05) <br><br> "Everybody wins, since those who are more fluent become role models to those who are learning." (P03) <br><br> "It stimulates those who don't speak the idiom to become better." (P02) | *Cooperation*: the use of tasks where the users must collaborate to achieve a common goal [22] (e.g., keep speaking a different idiom during the whole day). <br><br> *Social Pressure*: the social interactions exert pressure on the individual [22] (e.g., everybody is forced to keep speaking the idiom). | *Positive:* It creates learning opportunities, which helps increasing individual motivation. <br><br> *Negative:* It is not suitable for onsite use. | This practice might not be feasible for office environments due to insecurities of professionals (e.g., being observed by professionals from outside the team) and the possibility of noisy conversations (e.g., people speaking loudly might disturb other individuals at the office). |
| Project Happy Hour | "They are helpful moments for our mental health. Often these moments work as therapy for me". (P05) <br><br> "It helps to maintain a harmonious and pleasant relationship within the team". (P04) <br><br> "It makes people feel more connected, increasing the team bonding". (P03) | *Storytelling*: elements are added to support a narrative telling the story of the environment [22] (e.g., discussions about work were discouraged, while team members acted like they were in another environment, such a bar, or a casino). | *Positive:* It supports team cohesion and reduce individual isolation. <br><br> *Negative:* Not observed. | This practice could be used by software companies to deal with turnovers that are expected to occur in the period post-pandemic as it helps to maintain networking among professionals. |

## 5  Conclusions and Future Work

In this study, we present and discuss three practices that were used during the COVID-19 pandemic to support software teams while working from home, namely, *costume meeting*, *second language day* and *project happy hour*. We observed that these practices include some aspects of gamification—they all have explicit goals and rules and were created to produce fun and relaxation for team members.

These practices appear effective in increasing team engagement, team cohesion and reducing work monotony. Moreover, working from home during a crisis is different from working from home in normal conditions [4]. Therefore, we highlight the perspective of software professionals and the applicability of these practices in support to wellbeing. Furthermore, our study has implications for the use of ethnographies in software engineering, since we demonstrated the efficacy of this method to investigate human aspects of software engineering in the context of remote work.

Our research approach does not support statistical generalization from a sample to a population. Our findings may be *transferable*, in the sense that practitioners in other contexts may find these practices useful. Similarly, this kind of study cannot provide objective evidence of causality; rather, we can only say that the practices appear effective to us and to the participants. Moreover, given space limitations, our capacity for thick description is limited.

This study is part of a broader, ongoing investigation of the effects of the pandemic on software development, and how crisis-induced problems can be mitigated. The results presented in this paper stand alone in their presentation of three specific practices and use of theories of gamification [29][30] and social cohesion [33], to refine these practices and improve them in a way that they can support remote and hybrid teams.